\documentstyle[psfig,dina4]{article}

\begin{document}
\bibliographystyle{unsrt}

\vspace{-1.0truecm}
\begin{center}
\Large{ Impact Ionization Rate in ZnS }
\vskip 1.0truecm

\large{ Martin Reigrotzki, Michael Stobbe, Ronald Redmer }
\vskip 0.5truecm

\normalsize{ Universit\"at Rostock, Fachbereich Physik, \\
Universit\"atsplatz 3, D-18051 Rostock, Germany }
\vskip 0.5truecm

\large{ Wolfgang Schattke }
\vskip 0.5truecm

\normalsize{ Institut f\"ur Theoretische Physik,
Christian-Albrechts-Universit\"at Kiel,\\
Leibnizstra\ss e 15, D-24118 Kiel, Germany }
\vskip 1.0truecm

received:
\end{center}

\begin{abstract}
The impact ionization rate and its orientation dependence in {\bf k}
space is calculated for ZnS. The numerical results indicate a strong
correlation to the band structure. The use of a q-dependent screening
function for the Coulomb interaction between conduction and valence
electrons is found to be essential. A simple fit formula is presented
for easy calculation of the energy dependent transition rate.
\end{abstract}
\vskip 0.5truecm

\centerline{ PACS-No. 72.20.Ht, 72.20.Jv }

\normalsize

The impact ionization rate is of basic interest in understanding the
electron transport in semiconductors at high electric field
strengths. This microscopic quantity is necessary for calculating the
probability of the electron to ionize which is described by the
ionization coefficient $\alpha$.  Interesting features are the
orientation dependence and the threshold behavior, which were studied
for Si \cite{bh92,sy92,srs94,sy94} and GaAs \cite{srs94,wb92}.

Recently, we have determined the ionization rate for GaAs and Si
\cite{srs94}, following a scheme presented by Sano and Yoshii
\cite{sy92} for the evaluation of the impact ionization rate for Si
using a very efficient deterministic numerical procedure. Their
special integration scheme takes into account a rather high number of
points in the Brillouin zone so that they were able to study effects
of the anisotropy in {\bf k} space as well as the threshold behavior
for Si. Both were found to be direct consequences of the band
structure.

In this paper we investigate the impact ionization rate and its
orientation dependence in the wide band gap semiconductor ZnS
($E_G=3.63$ eV). The calculation is performed similar to \cite{srs94}
utilizing the integration method of Sano and Yoshii \cite{sy92} again.

Inserting a band structure which is obtained from local
pseudopotential parameters for ZnS, the correlation of the particular
band shapes with the impact ionization rate is shown.  Exchange and
umklapp processes are included in the evaluation of the impact
ionization rate. The consideration of an appropriate wave-vector
dependent screening function for the Coulomb interaction in
semiconductors is of significant influence on the results.
\vskip 0.45truecm


The impact ionization rate is usually determined from Fermi's
golden rule,
\begin{eqnarray} \label{1}
r({\bf k}_1,\nu_1)=
\frac{2\pi}{\hbar}\frac{\Omega^3}{(2\pi)^9}
\sum_{\nu_2\dots\nu_4} \int \dots \int
d^3 {\bf k}_2 d^3 {\bf k}_3 d^3 {\bf k}_4
\left|M_{tot}(1,2;3,4)\right|^2 \nonumber \\
\times \delta\left[ E_{\nu_1} ({\bf k}_1) + E_{\nu_2} ({\bf k}_2) -
E_{\nu_3} ({\bf k}_3) - E_{\nu_4} ({\bf k}_4) \right] \;,
\end{eqnarray}
where $\Omega$ and $\nu_i$ are the crystal volume and the
band index, resp. $M_{tot}$ denotes the matrix element for
ionization including direct (D), exchange (E) and umklapp processes,
\begin{equation} \label{4}
|M_{tot}|^2 = 2 |M_D|^2 + 2 |M_E|^2 - |M_D^{\ast} M_E + M_D
M_E^{\ast}| \,.
\end{equation}

The initial electron states belong to the conduction (1) and valence
band (2), whereas the final ones are both in the conduction band [(3)
and (4)].  The integrals are extended over the Brillouin zone.  The
ionization rate, Eq.\ (\ref{1}), is integrated over all directions in
{\bf k} space to get an averaged, only energy dependent impact
ionization rate
\begin{equation} \label{2}
R(E) = \frac{\sum\limits_{\nu_1} \int d^3 {\bf k}_1
        \: \: \delta\left[ E_{\nu_1}({\bf k}_1)-E \right]
        \: \: r({\bf k}_1,\nu_1)}
      {\sum\limits_{\nu_1} \int d^3 {\bf k}_1
        \: \: \delta\left[ E_{\nu_1}({\bf k}_1)-E \right] } \:.
\end{equation}

The analytical evaluation of the transition rate \cite{rob80,quad94}
requires strongly simplifying assumptions such as a parabolic band
structure and free electron wave functions, which are not well founded
for semiconductors with a direct (GaAs, ZnS) or even an indirect gap
(Si, ZnS when omitting the first conduction band which is not
contributing). Therefore, we numerically evaluate Eqs.~(\ref{1}) and
(\ref{2}) in order to take into account a realistic band structure and
corresponding wave functions in the entire Brillouin zone. This is
done using a grid method following Sano and Yoshii (Ref.~2), who made
extensive use of symmetry relations imposed by the crystal structure
and were able to take into account a great number of points in the
Brillouin zone for the evaluation of the integrations in Eq.~(\ref{1})
for Si. This method is accurate enough to treat even more sensitive
effects such as the threshold behavior or the orientation dependence
of the interband transition rate successfully.

We have calculated the band structure and wave functions within the
frame of the empirical pseudopotential method \cite{cb66}, utilizing
the set of pseudopotential parameters from Ref.~9. The expansion runs
over 113 {\bf G} vectors to produce a reasonable band structure in the
entire Brillouin zone and, especially, to give a realistic description
near the threshold. A plot of the band structure is presented in
Fig.~1.

The interaction between the electrons in the bands is described by a
statically screened Coulomb potential.  The consideration of the
wave-vector dependence of the dielectric function $\epsilon(q)$ is of
importance for the calculation of the matrix elements in
semiconductors as pointed out already by Laks et al.~\cite{lak+}.  For
the interaction between conduction and valence electrons, we apply a
q-dependent dielectric function derived by Levine and
Louie~\cite{ll}. This model incorporates the correct long-range and
short-range properties of the dielectric function and, in addition,
some consequences of the presence of a gap in the excitation spectrum
as characteristic of semiconductors.  The interaction between
electrons in the conduction band can be modeled by a Debye potential
with an inverse screening length $\kappa=[n_0 e^2/(\epsilon_0 k_B
T)]^{-1/2}$ for a temperature of $T = 300$ K and a corresponding
electron density of $n_0 = 10^{16}$ cm$^{-3}$. The choice of $n_0$ is
of less importance, as there are very few ionization events with
small momentum transfers of $q\approx \kappa$.

The {\bf k} vectors are restricted to the first Brillouin zone with
respect to the reduced zone scheme.  We want to mention at this point
that for umklapp processes we have to consider terms which go beyond
the expansion up to the 113 {\bf G} vectors of the pseudopotential
band structure calculation.  The respective expansion coefficients,
i.e.\ those belonging to reciprocal lattice vectors with $|{\bf G}| >
\sqrt{20} (\frac{2\pi}{a})$, were calculated by means of perturbation
theory~\cite{srs94,low51}.

For the numerical treatment of Eqs.~(\ref{1}) and (\ref{2}) the
$\delta$-function is replaced here by rectangles of unit area with a
height of $1/\delta E$ and a corresponding width of $\delta E$. A
value of $\delta E=0.1$~eV was found to be large enough to ensure good
convergence. This was proved by test calculations with a finer grid as
well as with a value of $\delta E=0.2$~eV. Of course, the broadening
factor is related to effects such as collision broadening and
intra-collisional field effect~\cite{bhi92}. However, the study of
such effects is not intended here.

An important feature of the grid method used in our previous paper
\cite{srs94} and in Ref.~2 is, that the storage of energy values and
wave functions can be restricted to the irreducible wedge (IW) if the
symmetry properties of the Brillouin zone are fully exploited.  The
wave function for a wave vector ${\rm T_k}{\bf k}$, which is obtained
by operating one of the 48 transformations ${\rm T_k}$ of the lattice
point group on a wave vector ${\bf k}$ of the IW, is then determined
from $\alpha_{\bf G}^{\nu}({\bf k})$ (see \cite{sy92,srs94}). The mesh
in {\bf k} space is constructed with regard to the limits of the IW,
\begin{eqnarray*}
0 &\le& k_x+k_y+k_z \le 1.5 (\frac{2 \pi}{a}) \:, \\
0 &\le& k_z \le k_y \le k_x \le (\frac{2 \pi}{a}) \:,
\end{eqnarray*}
as well as to the condition for the interval length of the assumed
grid $\Delta k_i=\frac{1}{n} (\frac{2 \pi}{a})$, $i=\{x,y,z\}$ with $n
\in \aleph$.

For the determination of the ionization rate $r({\bf k},\nu)$, we take
a mesh of 152 points in the IW [$\Delta k_i=\frac{1}{10}
(\frac{2\pi}{a})$]. For the study of the average ionization rate
$R(E)$, the summations over the conduction band indices were
restricted to the four lowest bands, whereas all four valence bands
were taken into account.
\vskip 0.45 truecm


In order to investigate the influence of the band structure (see
Fig.~1) on the ionization rate, we examine $r({\bf k}_1,\nu_1)$ [Eq.\
(\ref{1})], especially along three symmetry lines. In the first
conduction band there are no points at all satisfying the threshold
condition and, therefore, the total rate is due to electrons in the
higher -- mainly the third and fourth -- conduction bands. The
anisotropy of the ionization rate of ZnS for electrons initiating from
the second, third and fourth band is shown in Fig.~2.  Comparing these
results with the band structure data, a coincidence of the qualitative
behavior of the impact ionization rate for the three directions and
the third and fourth conduction bands can be stated. However, the
attempt to manifest this correlation by applying a power-law fit
formula for $r({\bf k}_1, \nu_1)$ as used in our prevoius paper
\cite{srs94} did not lead to a satisfactory result.  The validity of
this Keldysh-type fit \cite{kel} strongly relyes on the isotropy of
the matrix elements. Contrary to calculations for GaAs \cite{srs94}
and Si \cite{sy92,srs94}, these exhibit a notable dependence on the
{\bf k} value of the initial electron, which hence does not allow for
such a fit.

Furthermore, the necessity to consider the region of enlarged energies
above the second conduction band where band degeneracies and multiple
level crossings occur, complicates the correlation of impact
ionization to the band structure. For such high energies, also the
quality of empirical pseudopotential band structures becomes
questionable.  The desirable application of a nonlocal band structure
may lead to more rigorous results for the impact ionization rate.

The average energy dependent ionization rate $R(E)$ can, however, be
approximated by such a simple fit formula, which is given by
\begin{equation} \label{fit}
\tilde{R}(E) = P [ E-E_{th} ]^a \quad ,
\end{equation}
where the parameters are fixed to $P=5.14\times10^{10} s^{-1}
(eV)^{-a}$ and $a=5.183$. The threshold energy was set to
$E_{th}=3.8$~eV (The calculation gave first ionization events above
$E_{th}=3.93$~eV; regarding $E_{th}$ as a fit-parameter, best fits may
be obtained with $E_{th}\approx 3.5$~eV). Fig.~3 shows the result as
obtained from our numerical calculations together with the fit
$\tilde{R}(E)$. The increase of the curve indicates a threshold
behavior similar to that obtained for GaAs, a typical wide band-gap
semiconductor with a hard threshold.

\vskip 0.45 truecm


We have neglected in our present calculations phonon assisted impact
ionization processes as well as deep level ionization which may become
of importance at high fields.  Furthermore, other high-field effects
such as collision broadening or the intra-collisional field effect have
to be considered for a more comprehensive description of the transport
process of hot electrons in semiconductors similar to
\cite{bhi92,kra94}.

The aim of further calculations may be the study of the influence of
the obtained anisotropy effects on the ionization coefficient, where
an additional effect of the band structure is expected \cite{kim}.
Furthermore, the influence of dynamic screening which was found to be
substantial for Si \cite{yam} has to be studied also for ZnS. The
inclusion of the obtained impact ionization rate would also improve
present calculations for high-field transport in ZnS \cite{bre88,bgw93}.

\vskip 0.45 truecm


\centerline{ ACKNOWLEDGEMENTS }
\vskip 0.15 truecm

We would like to thank Steve Goodnick and Shankar Pennathur for
helpful comments.  This work was supported by the Deutsche
Forschungsgemeinschaft under grants Re 882/6-1 and Sha 360/8-1.


\vskip 1.0truecm


\newpage
\normalsize
\centerline{ FIGURE CAPTIONS }

\noindent
FIG.\ 1: Pseudopotential band structure of
ZnS. The labels indicate the conduction bands including degeneracies.\\
FIG.\ 2: Ionization rate for ZnS calculated for {\bf k} points along
three symmetry lines.\\
FIG.\ 3: Average ionization rate for ZnS obtained from the present
numerical calculation and from the fit formula (Eq.\ \ref{fit}),
indicating the contributions of the lowest four conduction bands.

\vskip 1.0truecm

\begin{figure}[h]
\centerline{\large FIG.\ 1}
\psfig{figure=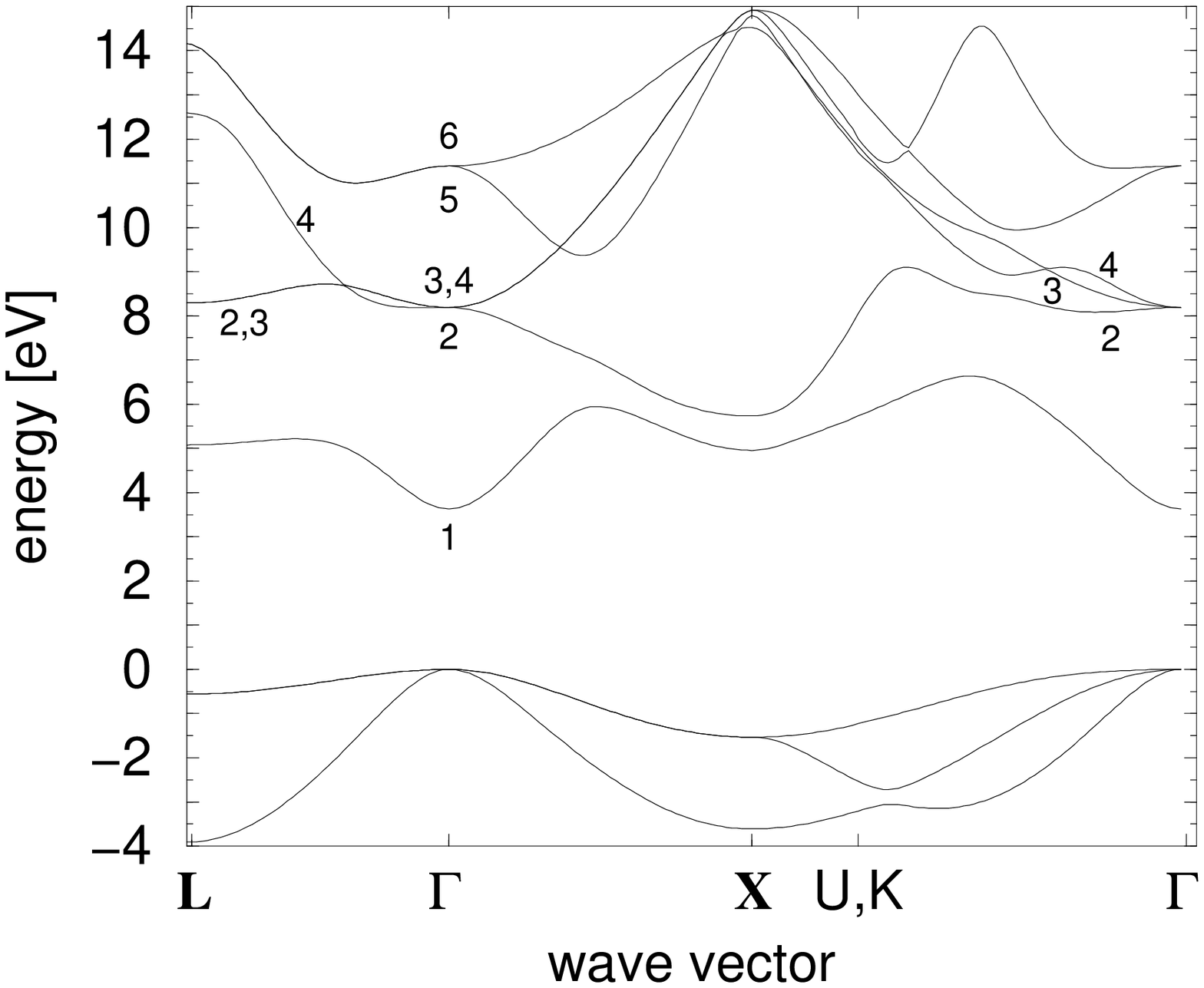,width=16cm}
\end{figure}
\begin{figure}[h]
\centerline{\large FIG.\ 2}
\psfig{figure=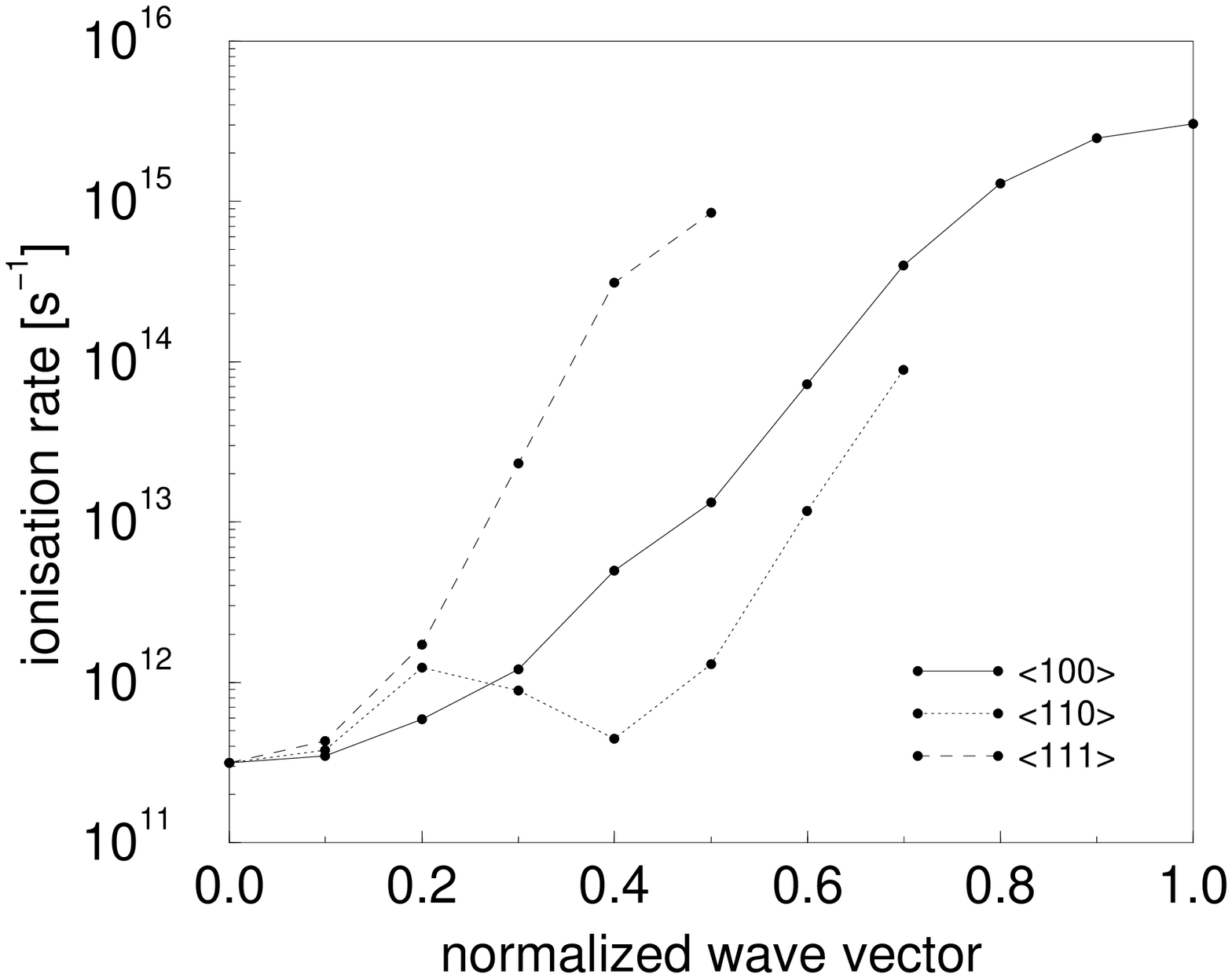,width=16cm}
\end{figure}
\begin{figure}[h]
\centerline{\large FIG.\ 3}
\psfig{figure=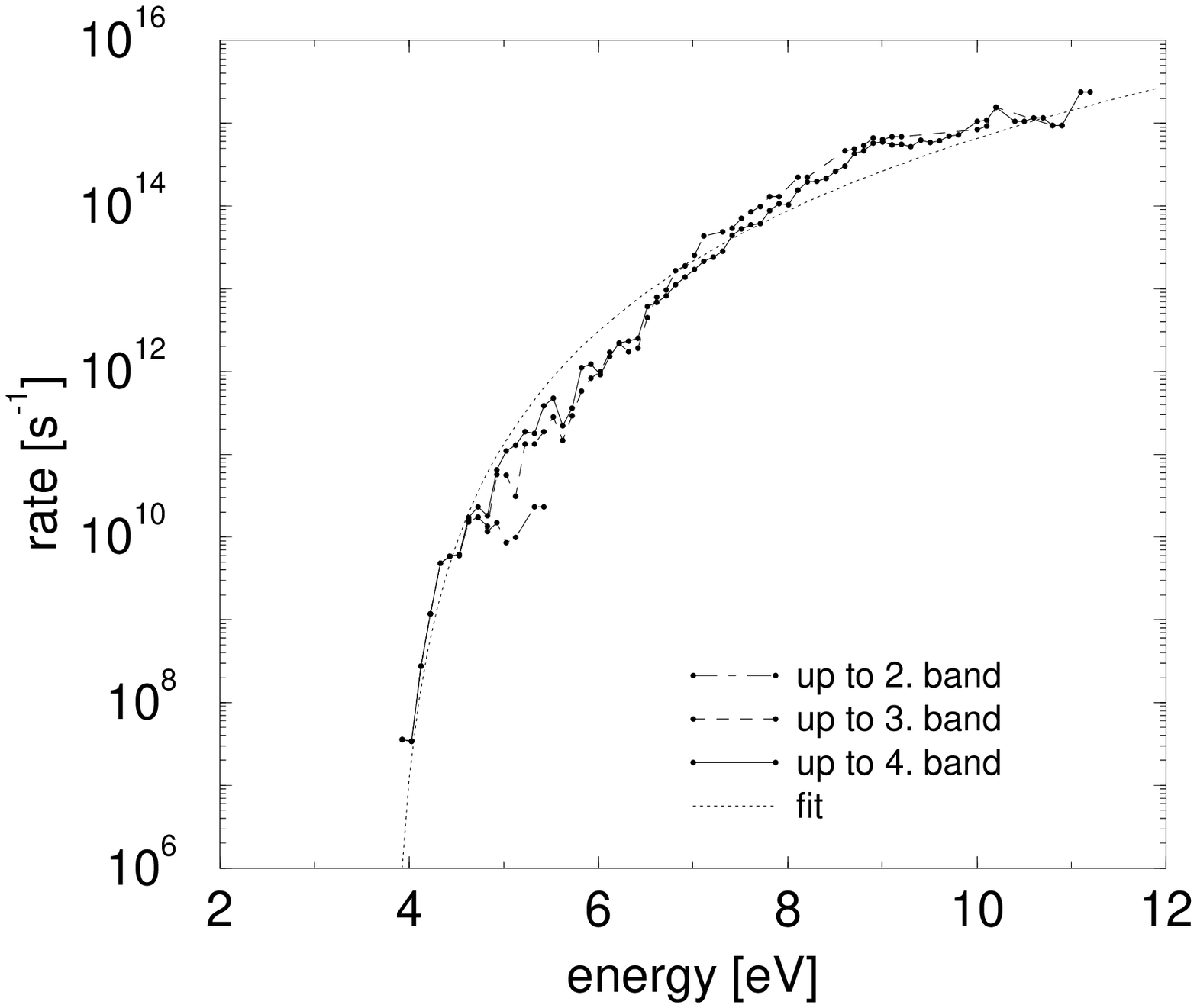,width=16cm}
\end{figure}

\end{document}